\documentclass[11pt]{article}

\usepackage{amsmath}
\usepackage{amssymb}
\usepackage{graphicx}
\usepackage{cite}
\usepackage{color} 

\usepackage[labelfont=bf,labelsep=period,justification=raggedright]{caption}

\date{}

\def\Ca{{\rm Ca}^{2+}}
\def\ca{${\rm Ca}^{2+}$}
\def\Cacon{[{\rm Ca}^{2+}]}
\def\cacon{$[{\rm Ca}^{2+}]$}

\def\caconin{$[{\rm Ca}^{2+}]_{\rm i}$}

\def\Caconj{[{\rm Ca}^{2+}]_{\rm NJ}}
\def\caconj{$[{\rm Ca}^{2+}]_{\rm NJ}$}

\def\ipt{IP$_3$}

\def\mum{\mu{\rm m}}

\def\M{{\rm M}}
\def\muM{\mu{\rm M}}
\def\mM{{\rm mM}}
\def\nM{{\rm nM}}

\def\bfig{\begin{figure}}
\def\endfig{\end{figure}}

\begin{document}

\begin{flushleft}
{\Large
\textbf{Cytoplasmic nanojunctions between lysosomes and sarcoplasmic reticulum 
are required for specific calcium signaling}
}
\\
Nicola Fameli$^{1,2,\ast}$, 
Oluseye A. Ogunbayo$^{2}$, 
Cornelis van Breemen$^{1,\dag}$
A. Mark Evans$^{2,\dag}$, 
\\
\bf{1} Department of Anesthesiology, Pharmacology,
and Therapeutics, The University of British
Columbia, Vancouver, British Columbia, Canada
\\
\bf{2} Centre for Integrative Physiology, University of
Edinburgh, Edinburgh, United Kingdom
\\
$\ast$ E-mail: nicola.fameli@ubc.ca
\\
$\dag$ co-senior authors
\end{flushleft}

\abstract{Herein we demonstrate how nanojunctions between lysosomes and
sarcoplasmic reticulum {(L-SR junctions)}
serve to couple lysosomal activation to regenerative,
ryanodine receptor-mediated cellular \ca\ waves.
In pulmonary artery smooth muscle cells
(PASMCs) it has been proposed that nicotinic acid adenine dinucleotide phosphate
(NAADP) triggers increases in cytoplasmic \ca\ via L-SR junctions, in a manner
that requires initial \ca\ release from lysosomes and subsequent \ca-induced
\ca\ release (CICR) via ryanodine receptor (RyR) subtype 3 on the SR
membrane proximal to lysosomes. {L-SR junction membrane separation has
been estimated to be $< 400$ nm}
and thus beyond the resolution of light microscopy, which has
restricted detailed investigations of the junctional coupling process. The
present study utilizes standard and tomographic transmission electron
microscopy to provide a thorough ultrastructural characterization of the L-SR
junctions in PASMCs. We show that L-SR nanojunctions are prominent features
within these cells and estimate that {the junctional membrane
separation and extension} are
about 15 nm and 300 nm, respectively. Furthermore, we develop a quantitative
model of the L-SR junction using these measurements, prior kinetic and
specific \ca\ signal information as input data. Simulations of NAADP-dependent
junctional \ca\ transients demonstrate that the magnitude of these signals can
breach the threshold for CICR via RyR3. By correlation analysis of live cell
\ca\ signals and simulated \ca\ transients within L-SR junctions, we estimate
that ``trigger zones'' with a {60--100} junctions are required to confer a
signal of similar magnitude. This is compatible with the {130} lysosomes/cell
estimated from our ultrastructural observations. Most importantly, our model
shows that increasing the L-SR junctional width above 50 nm lowers the
magnitude of junctional \cacon\ such that there is a failure to breach the
threshold for CICR via RyR3. L-SR junctions are therefore 
a pre-requisite for efficient
\ca\ signal coupling and may contribute to cellular function in health and
disease.}
\vspace{10mm}


\section{Introduction}
The importance of cytoplasmic nanojunctions 
to cellular signaling and thus to the modulation of cell function was
recognised several decades ago \cite{breemen_1977, Breemen_1988}, 
henceforth the extent to which cellular
nanospaces may contribute to the regulation of cell function received little
attention. Nevertheless there is now a growing recognition of the widespread
occurrence and functional significance of cytoplasmic nanospaces in cells
from species across several kingdoms 
\cite{Malmersjo_2013, breemen_2013, Moore_2012,
Carrasco_2011, nanospace_biophysics_2010,
edinburgh_2011, IUPS_2013}.

In this respect, membrane-membrane junctions between
lysosomes and the sarcoplasmic reticulum (L-SR junctions) are of particular
interest; not least because they have been hypothesized to couple calcium signaling
between these two organelles \cite{kinnear_2004,
kinnear_2008}.

That L-SR junctions may play an important role in cellular \ca\ signaling was
uncovered through early studies on the \ca\ mobilizing messenger nicotinic
acid adenine dinucleotide phosphate (NAADP) \cite{Lee_1995}, 
which demonstrated that NAADP released \ca\ from a store
other than the sarco/endoplasmic reticulum (S/ER) \cite{Aarhus_1996}, 
that could then trigger further \ca\ release from the
S/ER by \ca-induced \ca\ release (CICR) \cite{Lee_2000, Cancela_1999,
Churchill_2001, boittin_2002}. A major advance in our
understanding was then provided by the demonstration that the NAADP-released \ca\
was from an acidic lysosome-related store \cite{Churchill_2002,
yamasaki_2004, kinnear_2004} in a manner that requires
two pore segment channel subtype 2 (TPC2) \cite{Calcraft_2009}. 
However, studies on pulmonary
arterial smooth muscle cells (PASMCs) had also identified a significant
specialization, namely L-SR nanojunctions. It was hypothesized not only that
these nanojunctions were necessary for coupling between lysosomes and the SR
but that they could both coordinate and restrict their relationship to the SR
by preferentially targeting ryanodine receptors while excluding inositol 
1,4,5-trisphosphate (\ipt)
receptors \cite{kinnear_2004, kinnear_2008}.  
Importantly, NAADP-dependent \ca\ bursts primarily arise from
lysosomes in the perinuclear region of PASMCs and appear to promote rapid,
local \ca\ transients that are of sufficient size to activate clusters of SR
resident ryanodine receptor subtype 3 (RyR3) and thus initiate, in an
all-or-none manner, a propagating global \ca\ wave \cite{kinnear_2004,
kinnear_2008}.

The specialization of the proposed L-SR junction is likely important
in smooth muscle cell physiology, e.g., in vasomotion, and in the recycling
of organelles and programmed cell death by autophagy
\cite{Schneider_1967}. However,
L-SR junctions may also make as yet unforeseen contributions to
vascular pathologies as highlighted by the fact that Niemann-Pick
disease type C1 results, in part, from dysregulation of lysosomal
\ca\ handling \cite{lloyd-evans_2008} 
and is known to precipitate portal hypertension \cite{Tassoni_1991}, 
while other lysosomal storage diseases (e.g., Pompe and Gaucher
disease) accelerate pulmonary arterial hypertension
\cite{Noori_2002, Jmoudiak_2005}.
Moreover, observed hypertension is often associated with dysfunction
of cholesterol trafficking \cite{Carstea_1997}, 
increased plasma cholesterol levels,
vascular lesion formation, atherosclerosis/thrombosis and medial
degradation \cite{Tassoni_1991, Ron_2008}.
Therefore lysosomal \ca\ signaling is of considerable clinical interest.
That L-SR junctions may be of further significance to pathology is also
evident, for example, from the fact that in the pulmonary artery smooth muscle
L-SR junctions underpin \ca\ waves
initiated by endothelin 1, the levels of which are elevated in pulmonary 
hypertension, systemic hypertension
and atherosclerosis \cite{Shao_2011, Davie_2009}. 
An understanding of how 
specific \ca\ signals are functionally initiated 
therefore has important translational implications.

Lysosomal \ca\ regulation has been of
current interest in several recent research and review articles
(e.g., \cite{Wang_2012, 
Cang_2013, Morgan_2013, kilpatrick_2012, Collins_2011}), 
yet the mechanism by which \ca\ signals are  generated by the
endolysosomal system has not yet been modeled in a truly quantitative manner.
Given the likely importance of L-SR junctions to \ca\ signaling in 
health and disease, we sought to determine whether this nano-environment 
would indeed be able to
effectively generate the previously observed NAADP-induced \ca\ signals. 

Due to the minute spatial
scale of the nanojunctions generating the primary \ca\ signals, 
accurate investigation of dynamic signaling within these 
spaces cannot be achieved with currently available
instrumentation. 
To overcome this issue, we took an integrative approach by
combining our own electron microscopy of L-SR nanojunctions,
existing kinetic data on the \ca\ transporters and buffers, 
and prior knowledge of the 
NAADP-induced \ca\ signal features into a 
quantitative model of a typical L-SR nanojunction.
This model is based on stochastic simulations of intracellular 
\ca\ diffusion by Brownian motion implemented using the particle simulator
MCell  (freely available at mcell.org)
\cite{stiles_1996, stiles_2001, kerr_2008}. 
In particular, we set out to verify the following hypotheses in PASMCs:
(1) L-SR nanojunctions should be observable in the ultrastructure of
these cells, 
(2) NAADP induces sufficient \ca\ release from the lysosome to initiate
activation of RyR3 embedded in the junctional SR, and (3) the combined
effect of activation of L-SR nanojunctions in a cytoplasmic ``trigger zone''
determines the threshold of global \caconin\ for the biphasic release
process. 

In the present manuscript we have verified the existence of L-SR nanojunctions
within the ultrastructure of PASMCs, and
shown that lysosomes can release sufficient \ca\ to activate CICR
via RyR3 clusters embedded in the junctional SR. Perhaps most
importantly, we show that L-SR coupling is determined both by the
integrity of L-SR junction on the nanoscale and the quantal summation
of \ca\ release from multiple, activated junctional complexes.

\section{Results}
\subsection{NAADP-induced $\boldsymbol{\Ca}$ signals within isolated 
pulmonary artery smooth muscle cells}
{The relevant background
findings that stimulated the development of the work presented here
were first reported by Evans' group in \cite{boittin_2002} and
\cite{kinnear_2004}, and are summarized in figure \ref{Fig_1}.} 
\bfig[!ht]
\centering
\includegraphics*[scale=.5]{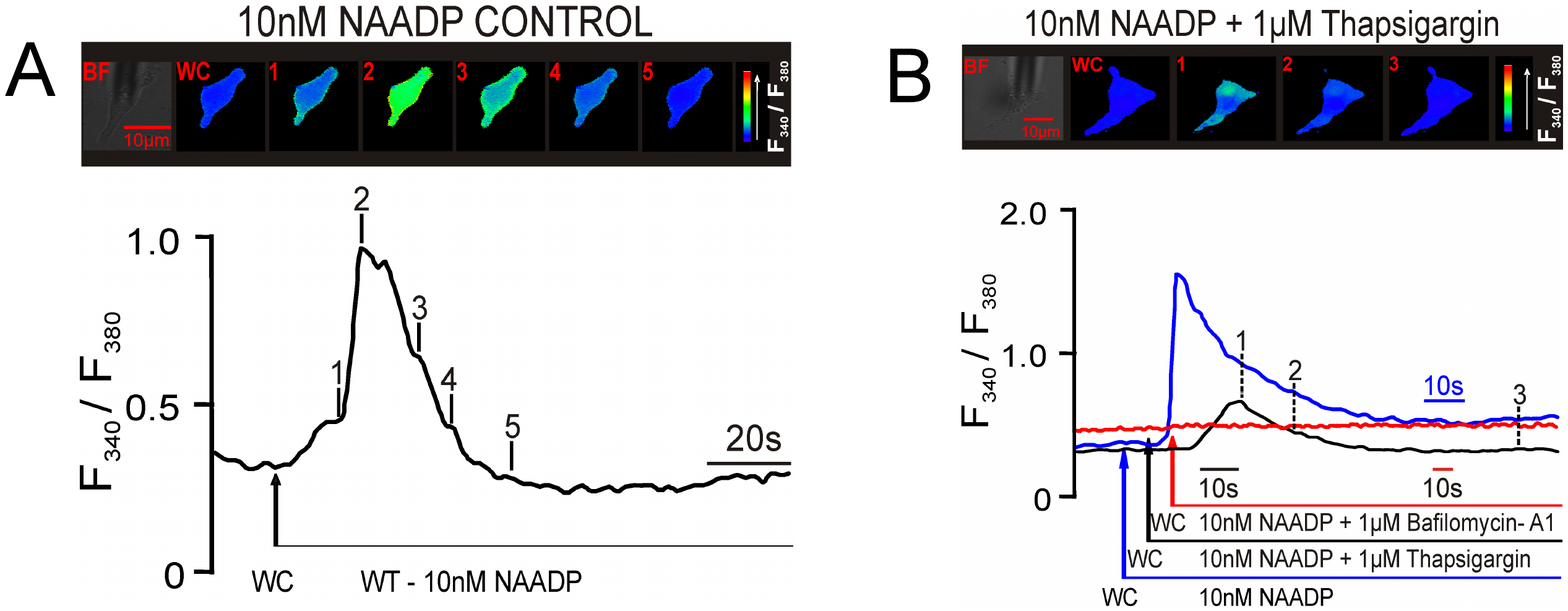}
\caption{{\textbf{
A, B,} Upper panels show a series of pseudocolour images of the Fura-2
fluorescence ratio (F340/F380) recorded in two different pulmonary artery
smooth muscle cell during intracellular dialysis of 10nM NAADP before (A) and
after (B) depletion of SR stores by pre-incubation (30 min) with
$1\;\muM$
thapsigargin. Note the spatially localized `\ca\ bursts'. A, Lower panel shows
the record of the Fura-2 fluorescence ratio against time corresponding to the
upper left panel of pseudocolours images; note the discrete shoulder in the
rising phase of the F340/F380 ratio that corresponds to the initial `\ca\
bursts'. B, Lower panel shows paired responses to 10 nM NAADP under control
conditions (black), following depletion of SR \ca\ stores with $1\;\muM$
thapsigargin (blue) and following depletion of acidic stores with $1\;\muM$
bafilomycin-A1 (red). Scale bars: $10\;\mum$.}
}\label{Fig_1}
\endfig

{
The example record in figure \ref{Fig_1}A 
highlights the fact that NAADP appears to activate a
two-phase \ca\ signal, which can exhibit an identifiable ``shoulder'' during the
initial rising phase of the signal (figure \ref{Fig_1}A, time point 1), followed by a
second faster phase of signal amplification (figure \ref{Fig_1}A, time point 2). It is
notable that the delay to the initiation of the second phase of amplification
is variable \cite{boittin_2002, kinnear_2004, kinnear_2008}
and due to this fact the shoulder is not always evident (see for example
figure \ref{Fig_1}B, lower panel). Previous studies have demonstrated that this
two-phase response results from initiation by NAADP of \ca\ bursts from
lysosome-related stores \cite{kinnear_2004}  in a manner that requires TPC2
\cite{agbani_2011} and that \ca\ bursts  are subsequently amplified by CICR
from the SR via clusters of RyR3  \cite{boittin_2002, kinnear_2008}.
Panel B of this figure illustrates the fact that prior depletion of SR stores
by pre-incubation with thapsigargin ($1\;\muM$) blocks the amplification phase,
while depletion of acidic \ca\ stores with bafilomycin abolishes the entire
NAADP-induced \ca\ signal. The previous studies cited above provided
calibrated estimates of the changes in intracellular \cacon\ input into the
model below.}

Based on 
earlier optical microscopy work, like the data in figure
\ref{Fig_1}, and immunofluorescence results, 
it has been proposed that, for the lysosomal \ca\ release to trigger CICR,
L-SR nanojunctions are required and that they
consist of apposing
patches of lysosomal and SR membranes separated by a narrow space of
nano-scale dimension \cite{kinnear_2004, boittin_2002}.
These studies led to an upper limit of 400 nm {for the separation of the
junctional membranes.}
We propose that L-SR junctions do indeed 
represent cellular nanojunctions and 
that they might play a role of accentuating \ca\ gradients, akin to
that of {plasma membrane (PM)}-SR 
junctions that are pivotal in the process of SR \ca\ refilling
 during asynchronous \cacon\ waves 
\cite{fameli_2007}.
We furthermore hypothesize that in order for these nanojunctions 
to appropriately
regulate \ca\ signaling, they must be separated by a distance of
approximately 20 nm
and be of the order of a few hundred nm in lateral dimensions, as inferred
from previous studies on PM-SR junctions \cite{kinnear_2004, kinnear_2008}. 
\subsection{Ultrastructural characterization of L-SR nanojunctions}
To identify lysosomes, SR regions and L-SR nanojunctions, we recorded
and surveyed
74 electron micrographs of rat pulmonary arterial smooth
muscle taken from samples prepared as described in the Materials and Methods 
section.
The images in figure \ref{Fig_2} provide a set of examples. 
{Since we were aiming to detect L-SR junctions, namely close appositions of
the lysosomal and SR membranes, immuno-gold
labeling of lysosomes was prohibited, since this technique compromises
membranes definition by electron microscopy to the extent that we would be
unable to assess junctional architecture. Instead, in images
like those in figure \ref{Fig_2}, 
lysosome identification was accomplished by relying on the 
knowledge of lysosomal
ultrastructural features, which has accumulated over the past 50 years since
the discovery of the lysosomes (see for example, \cite{histology_74}).}
\bfig[!ht]
\begin{center}
\includegraphics*[scale=.55, viewport=14 120 590 701]{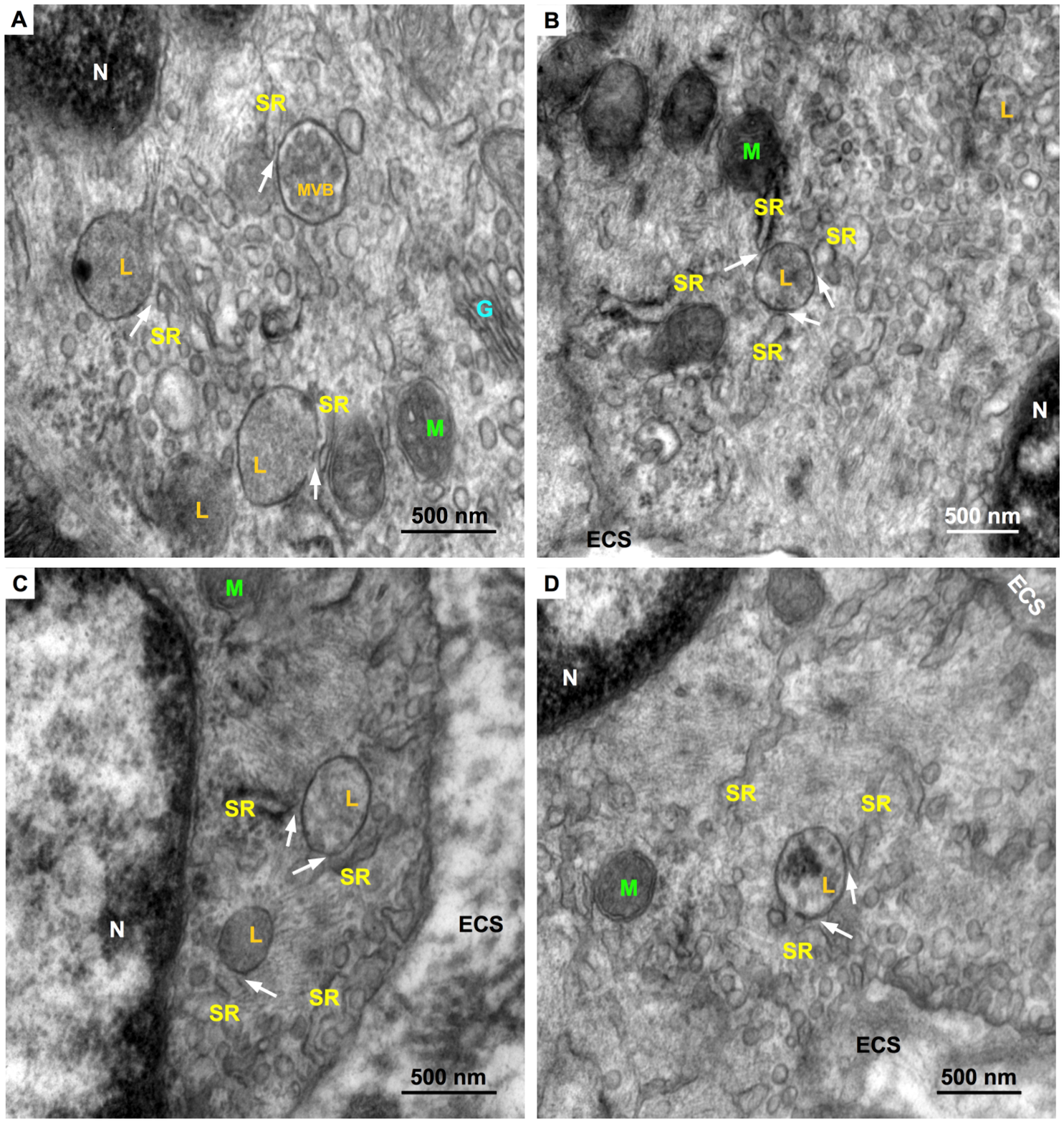}
\end{center}
\caption{{Representative electron
micrographs of rat pulmonary artery SMC regions
containing lysosomes (L), several SR cisterns, and including several
examples of L-SR junctions (arrows). Also indicated are nuclei
(N), Golgi apparatus (G), mitochondria (M), {a multivesicular body
(MVB)} and extra-cellular space (ECS).}
${\rm Scale\; bars}=500$ nm.
Magnifications: A,C 80,000$\times$, B,
60,000$\times$, D, 70,000$\times$.}\label{Fig_2}
\endfig

In standard (2D) transmission electron microscopy (TEM) images, lysosomes
typically appear as elliptical structures bound by a single lipid
bilayer, a feature that distinguishes them from mitochondria. 
{Depending on the lysosomal system stage, they also
tend to have a more or less uniformly electron-dense 
interior as compared to the surrounding 
cytosol \cite{Robinson_1986}. 
They can be distinguished from endosomes by their larger
size and darker lumen and they differ from peroxisomes,
since the latter usually display a geometrically distinct and markedly darker
structure called ``crystalloid'' in their lumen. Moreover, it would appear that
peroxisomes are found far more frequently in liver, kidney, bronchioles and
odontoblasts than in other cell types (see, for example, \cite{histology_74,
Ovalle_2013, em_atlas}).
Occasionally, organellar remnants are still visible inside these
ovals, a characteristic that identifies them as multi-vesicular
bodies (``MVB'' in figure \ref{Fig_2}A). As it is at times questionable
whether MVB's are late endosomes or endosome-lysosome hybrids (compare, for
example, \cite{histology_74} and \cite{holtzman_1989}), we have excluded
organelles (3 in total) with such characteristic
from our statistical count.}

From each of the relevant smooth muscle regions surveyed, we obtained
high-resolution images of areas containing lysosomes and 
L-SR junctions in order to
quantitatively characterize them (arrows in figures \ref{Fig_2}
and \ref{Fig_3}A). 
Using a software graphics editor
(inkscape.org) and the image scale bar as a calibration gauge, 
we measured the lysosome size, as the length
of the major and minor axes of their elliptical 2D projections (in
orange and grey, respectively, in figure \ref{Fig_3}A), the
L-SR widths, that is the distance
between lysosomal and SR membranes at places where the two were about
30 nm or closer to each other (in purple in figure \ref{Fig_3}A), 
and the L-SR extensions as a percentage ratio between the junctional
SR and the lysosomal membranes (in turquoise in figure \ref{Fig_3}A).
From these measurements, we extrapolated the 3D junctional SR extension, both as a
percentage of the lysosomal surface and as a length in nm.
The histograms displayed in figure \ref{Fig_3}B--D 
visually summarize the data
collected from the image analysis. The mean and standard deviation
values of the measured parameters are reported in table \ref{l_sr_params}.
\bfig[!ht]
\begin{center}
\includegraphics*[scale=.4]{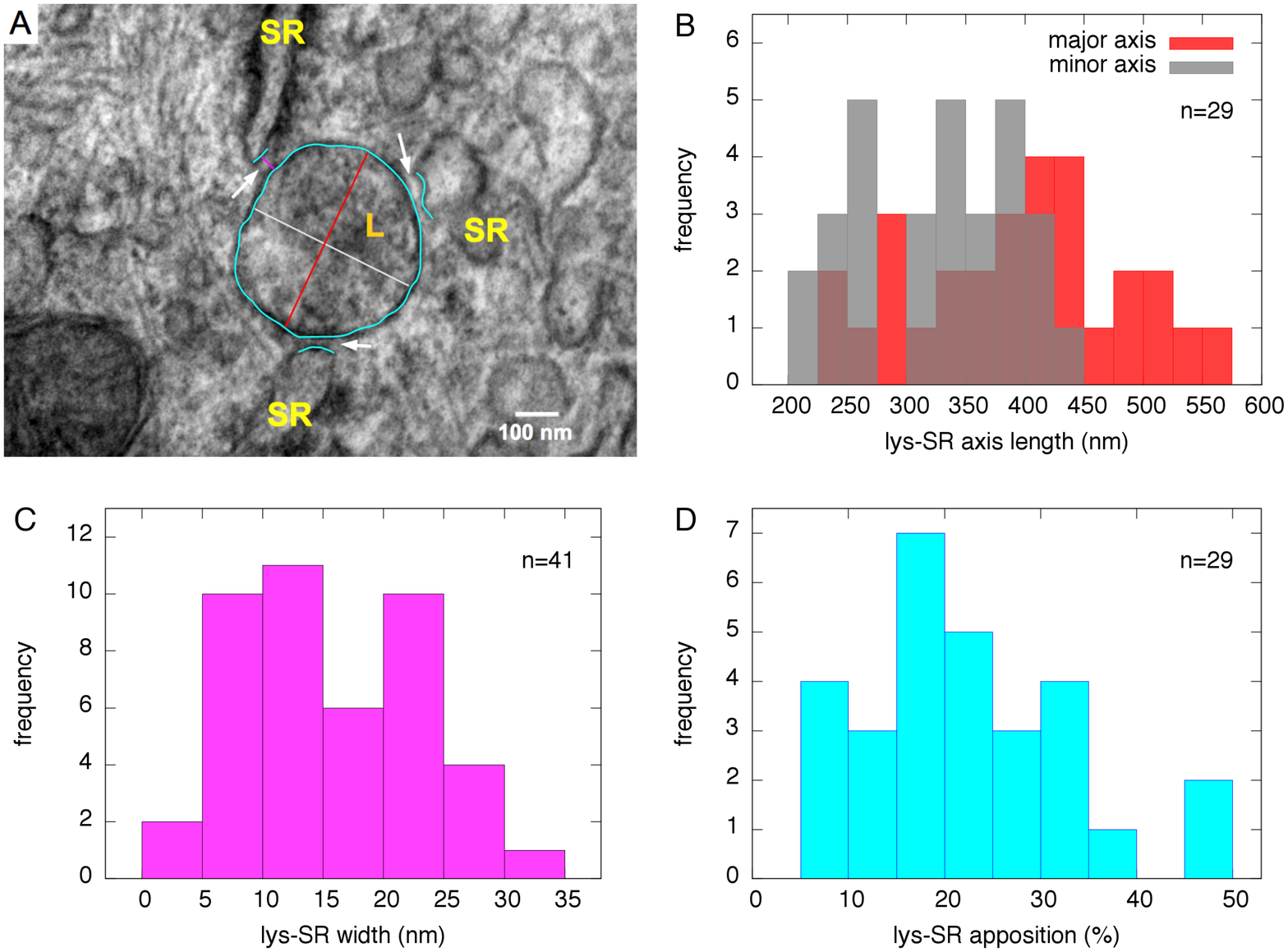}
\end{center}
\caption{\textbf{A,} High magnification (150,000$\times$)
electron micrograph of  a region of Figure \ref{Fig_2}B containing
3 L-SR junctions (arrows);
coloured tracings as shown were
used to measure lysosome dimensions, L-SR widths and extensions.
 ${\rm Scale\; bar}=100$ nm.
 \textbf{B--D,} Histograms showing distribution of several relevant
 lysosomal and L-SR junctional parameters, used to characterize the
 junctions and inform the quantitative model. B, lysosomal
 dimensions as major and
 minor axes of oval shape in micrographs; C, L-SR junctional width; D,
 percentage {apposition} between junctional SR and lysosome
 perimeter as projected in 2D micrographs.
 }\label{Fig_3}
 \endfig

Estimates of the various parameters gathered in this phase of the study 
were used as a basis to build a 3D software object to
represent a typical L-SR junction. This reconstruction was then used to
design the simulations mimicking \ca\ diffusion in the L-SR
nanojunctions, as is described below.
\begin{table}
\begin{center}
\caption{
\bf{L-SR junction characterization parameters}}\label{l_sr_params}
        \begin{tabular}{c|c|c}\hline\hline
        parameter & mean $\pm$ SD & notes\\\hline
        lysosome minor axis & $(325\pm 65)$ nm & $n=29$\\
        lysosome major axis & $(398\pm 91)$ nm & $n=29$\\
        L-SR width & $(16\pm 7)$ nm & $n=41$\\
        L-SR overlap & $(23\pm 13)$ \% & $n=29$\\
        L-SR lateral extension & $(262\pm 209)$ nm & calculated from
data above\\
        3D L-SR overlap & $(13\pm 15)$ \% & extrapolated from 2D
measurements\\\hline\hline
        \end{tabular}
\begin{flushleft}
Mean and standard deviation values of L-SR junction parameters,
calculated from data as in Figure \ref{Fig_3}B--D.
\end{flushleft}
\end{center}
\end{table}

\subsubsection{Tomography}
To gather more direct information on the 3D morphology of L-SR junctions, we
acquired a set of tomograms of those regions from the same sample
blocks used to obtain the images in figure \ref{Fig_2}. 
In figure~\ref{Fig_4}, we report snapshots from one of the tomograms; 
in these stills, we have also traced out parts of one
lysosome and the closely apposed SR region that together form a L-SR junction.
These tomograms are very helpful in clarifying the detailed morphology of
L-SR junctions and informing on the spatial
variability of the SR network. For example, while the SR segment shown 
in a single 2D tomographic scan (figure
\ref{Fig_4}A) would appear to be continuous and part of a large SR
compartment, it 
actually branches out into narrower cisterns as revealed by 3D tomographic 
reconstruction (figure \ref{Fig_4}B). 
Furthermore, it is interesting to observe 
the fact that one extension
of the SR appears to couple with multiple {organelles}. Thus, the
3D views generated by tomography are paramount for demonstrating the
presence of a true junctional complex and for the design of a
prototypical L-SR environment as a software mesh object, on which we may
simulate the NAADP-mediated localized \ca\ release.
\bfig[!ht]
\begin{center}
\includegraphics*[scale=.5, viewport=0 150 822 455]{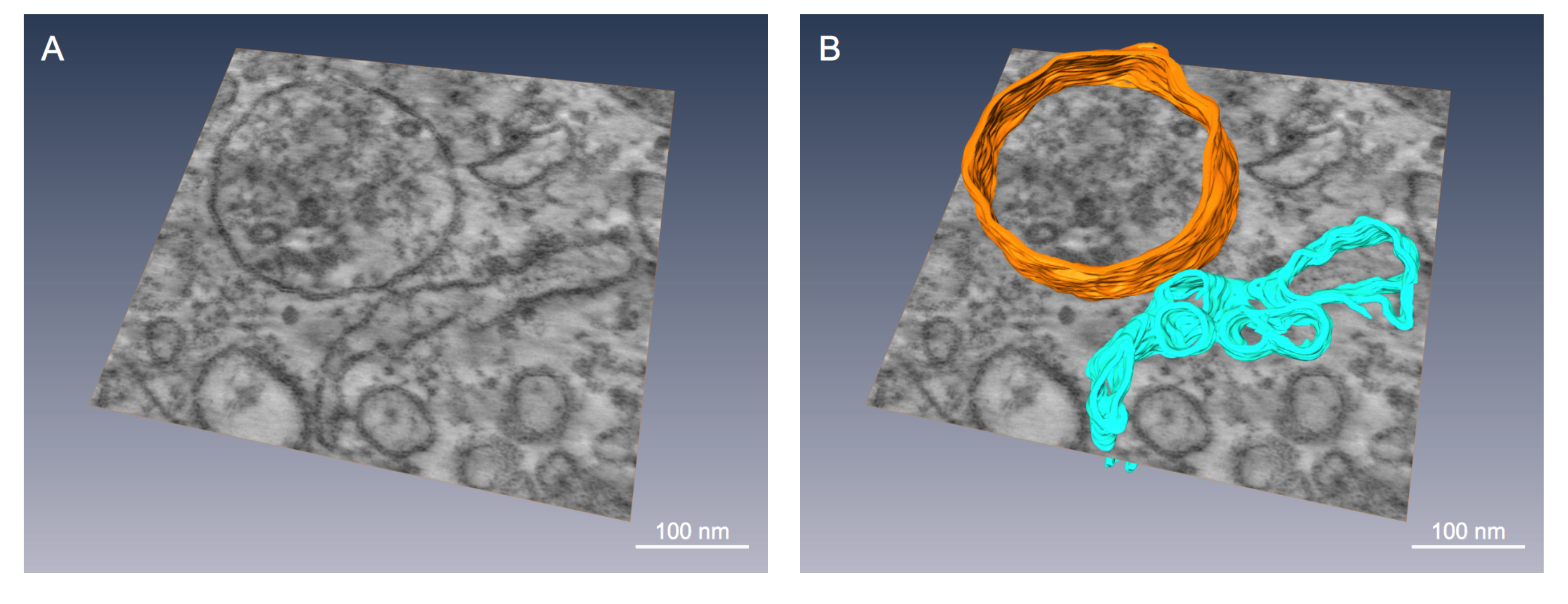}
\end{center}
\caption{\textbf{A,} Snapshot from a TEM tomogram of
a L-SR region of rat pulmonary artery smooth muscle, illustrating, among
other things, a single SR extension apparently forming junctions with
several lysosomes. 
${\rm Magnification}=62,000\times$.
\textbf{B,} Same snapshot shown in A, but with a lysosome (orange) and
a portion of SR (turquoise) partially traced out in 3D.
These pseudo-colour tracings
underscore how the SR can appear as a large cistern in a given plane, but
can
actually branch out in different directions when viewed in 3D.
Scale bars $\approx 100$ nm.}\label{Fig_4}
\endfig

\subsection{Quantitative model} 
The model aims to verify whether 
NAADP-induced \ca\ release from the lysosomal system could be
responsible for the localized \ca\ signal preceding the global \ca\
wave (see figure \ref{Fig_1} and
\cite{kinnear_2004}), which triggers a propagating wave by CICR via RyR3s
localized at L-SR junctions, as predicted by earlier
observations \cite{kinnear_2004}. 
\subsubsection{Generation of the signal's ``shoulder'' by lysosomal
$\boldsymbol{\Ca}$ bursts}
To understand the generation of the signal ``shoulder'', 
such as that
observed at time point 1 in the lower panel of 
figure \ref{Fig_1}A, by \ca\ bursts within L-SR nanojunctions, let us note that 
its magnitude corresponds to the difference in \cacon\ between the 
resting level of $\approx 100$ nM \cite{boittin_2002}
prior to NAADP stimulation (up to the point, at which NAADP enters the
cytoplasm under the whole-cell configuration (WC) in figure
\ref{Fig_1}) and the value of
$\approx 400$ nM \cite{boittin_2002}, 
corresponding to \caconin\ at time point 1 in the example record shown in figure
\ref{Fig_1}A, lower panel {(these concentration values were obtained via a
standard calibration procedure \cite{boittin_2002})}. 
This leads to a
$\Delta\Cacon_{\rm shoulder}$ of approximately 300~nM \cite{boittin_2002}. 
Let us now estimate the number of lysosomes required to generate
a $\Delta\Cacon_{\rm shoulder}$ of such magnitude in a PASMC,
given the following assumptions:
\begin{enumerate}
\item That the luminal \cacon\ of a lysosome, $\Cacon_{\rm lys}$, is
{in the range of $400$--$600\;\muM$}, as determined in mouse macrophages
\cite{christensen_2002} {and that it is homogeneous across the lysosomal
population};
\item That the \ca\ release rate {during bursts is $\approx 10^6$
ions/s (based on findings in \cite{Pitt_2010}, 
but see next section for a detailed analysis on the rate
time-variation) and that it}
becomes negligible for values of $\Cacon_{\rm lys}$ below $\approx 80\;\muM$, as
suggested by the single channel kinetics of TPC2 signaling complexes
in lipid bilayer studies 
\cite{Pitt_2010};
\item That lysosomes are spheres with a radius of
$\approx 180$ nm, as gathered from our EM characterization 
(see figures \ref{Fig_2}, \ref{Fig_3}, \ref{Fig_4}, and table
\ref{l_sr_params}){, and hence that their volume is $V_{\rm
lys}=(4\pi/3)(1.8\times 10^{-7}\,{\rm m})^3=2.4\times 10^{-17}\,{\rm L}$}; 
\item That a  smooth muscle cell cytosolic volume can be calculated as 
{$V_{\rm
cyt}=2.4\times 10^{-12}$ L} by
modeling a cell as a 130-$\mum$-long
cylinder, 6 $\mum$ in diameter \cite{Somlyo_1975}, and accounting for 
nuclear,
SR, mitochondrial and lysosomal volumes.
\end{enumerate}
With these assumptions accepted, {
from points 1. and 2. above we gather that
a lysosome can release a potential 
$\Delta\Cacon_{\rm lys}=(400\;{\rm to}\;600)\muM-80\muM=3.2\times
10^{-4}$~M to $5.2\times 10^{-4}$~M into the cytosol 
(as we elaborate in the next section, this should take $\approx 0.03$~s, a 
time frame that ensures that released \ca\ can be considered unbuffered).
Each lysosome contribution to \caconin, $\Delta\Cacon_{\rm i,lys}$, 
can then simply be calculated by taking the lysosome-to-cell volume 
ratio into account:
\begin{equation}
\Delta\Cacon_{\rm i,lys}=\Delta\Cacon_{\rm lys}
\frac{V_{\rm lys}}{V_{\rm cyt}}\approx 3\;{\rm to}\;5\;\nM
\end{equation}}
{Therefore, we can calculate the number of lysosomes that may contribute to
the magnitude of $\Delta\Cacon_{\rm shoulder}$ as
\begin{equation}
\frac{\Delta\Cacon_{\rm shoulder}}{\Delta\Cacon_{\rm i,lys}}\approx\;60\;{\rm
to}\;100\;{\rm
lysosomes}\label{lys_4_shoulder}
\end{equation}}
In summary, {between 60 and 100} lysosomes would be necessary (and
possibly sufficient) to provide a $\Delta\Cacon_{\rm shoulder}$ of 300 nM, which is
typically observed during the localized \ca\ release phase of the NAADP-induced \ca\
signals.

How many lysosomes do we actually expect to be in a PASM cell of our sample
tissue? We can obtain a rough estimate of this number from the TEM imaging we
performed for this study. 
In each 80-nm-thick TEM sample section, we see {5--10} lysosomes/cell. 
Lysosomes are predominantly localized to
the perinuclear region of the cytoplasm, but seldom in the subplasmalemmal
area, consistent with previous observations by optical microscopy
\cite{kinnear_2004, kinnear_2008}. 
If we simplify the geometry of a typical smooth muscle cell to a
130-$\mum$-long cylinder with radius of 6 $\mum$, and
considering that lysosome radii are around 180 nm, as mentioned above,
it is reasonable to assume that
separate sets of lysosomes would be observable in TEM images 
taken at distances into the sample of about 180 nm from each other.
Neglecting the slices within one lysosomal diameter of the cylinder
surface---given the near total lack of observed lysosomes in those
subvolumes---then our images suggest that we can expect a total of about
{130}
lysosomes/cell.
It is encouraging that we obtain from this count a higher number than the 
{60--100 lysosome range} we derived in equation 
(\ref{lys_4_shoulder}), in that on the one hand it is plausible to
think that not all of the lysosomes in a cell may be activated in synchrony,
nor may they all be involved in NAADP-mediated 
signaling, and on the other hand
experience tells us that evolution has built in some
redundancy of function in order to provide a threshold and also a margin of
safety
for the generation of this type of \ca\ signals. Moreover,
the estimated number of junctions is based
on a value for $\Cacon_{\rm lys}$ determined in macrophages, and it is
plausible that the total
releasable $\Cacon_{\rm lys}$ in a PASMC may differ from that value
{and that it may also vary over the lysosome population}.
Lastly, the uncertainty in the number of lysosomes/cell evidenced from the electron
micrographs as described above may also contribute to this discrepancy.
\subsubsection{L-SR junctional $\boldsymbol{\Ca}$ signal}
In the hypothesized model outlined
in \cite{kinnear_2004, kinnear_2008},
\ca\ bursts activate SR resident RyR3 channels
within L-SR junctions and thus initiate a propagating \ca\ wave by CICR.
The stochastic simulations developed here
attempt to reproduce the phenomenon of the generation of \cacon\
transients within individual
L-SR junctions, considering the \ca\ release kinetic requirements for
lysosome-resident TPC2 signaling complexes and the rate of \ca\ capture
by the SERCA2a
localized on the neighbouring SR membrane \cite{clark_2010}, and to link these
junctional transients to the observed bursts.

The thorough quantitative image analysis described in the previous
section yields critical information for our first modeling phase, in
which we built a dimensionally accurate
virtual lysosome and a portion of the SR system, closely
apposing the lysosome (figure \ref{Fig_6}),
so as to reproduce a NAADP-triggered \ca\ signal
within the nanospace of a representative L-SR junction as faithfully as possible.
\bfig[!ht]
\begin{center}
\includegraphics*[scale=.45, viewport=0 34 836 605]{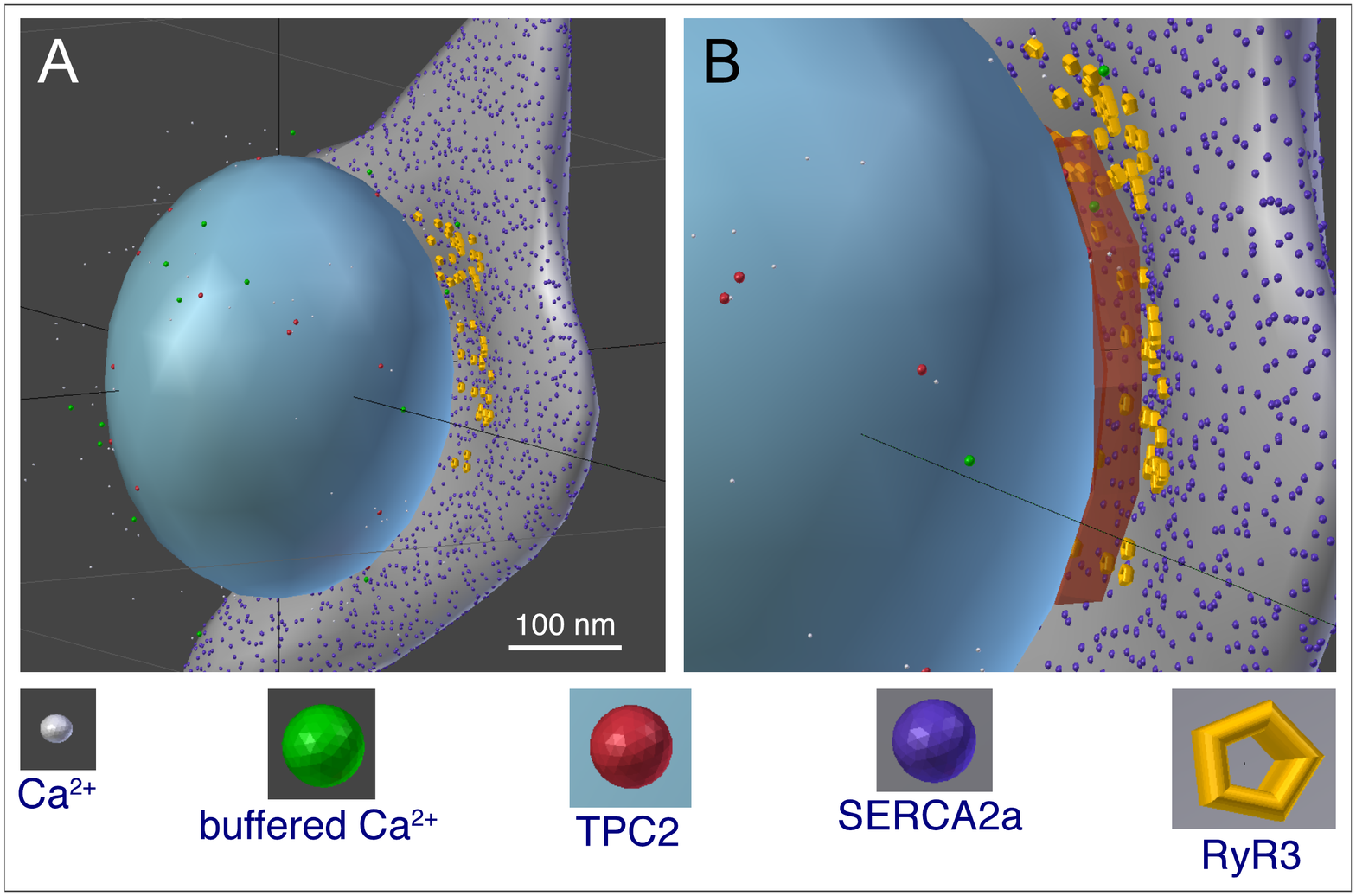}
\end{center}
\caption{\textbf{A and B,}
3D software reproduction of a lysosome closely apposed to a portion of SR,
thereby forming an $\approx\;20$-nm-wide
L-SR nanojunction; this rendering was inspired by a series
 of observations from micrographs as in
 Figure~\ref{Fig_2} (grey object$=$SR,
 blue object=lysosome). Included are relevant molecules
 traversing Brownian motion trajectories produced by the model
 simulations (see symbol legend below the panels).
 B, enlarged view of the L-SR junctional region, in
 which we have displayed the volume object {(rust-coloured box)}
 used to measure the $\Cacon_{\rm
 NJ}$ transients like the ones reported in figure~\ref{Fig_7}. {Scale
 bar~$\approx$~100~nm. The model
 geometry and code files are available from the corresponding author.}}\label{Fig_6}
 \endfig

From the available literature we estimated the number of
TPC2 and SERCA2a likely
distributed on the lysosome and SR membranes, respectively. 
We obtained the former number by dividing
the macroscopic whole-lysosome conductance, calculated 
from the current values reported
in a recent study on TPC2-mediated \ca\ current
in isolated lysosomes \cite{Schieder_2010a}, by the single channel
conductance determined in \cite{Pitt_2010}, thus estimating that a typical
lysosomal membrane  may contain $\approx 20$ TPC2.
To obtain this value, 
we used the experimental condition data provided in 
\cite{Schieder_2010a} to extract values for the ionic potential, $E_{\rm ion}$, 
across the lysosomal membrane. We then employed the authors' current-voltage
($I$-$V_{\rm m}$) data to compute whole-lysosome
conductance values ($g_{\rm WL}$) as a function of the applied membrane potential, 
$V_{\rm m}$, from Ohm's law: $I=g_{\rm WL}(V_{\rm m}-E_{\rm ion})$. 

Moreover, we have estimated the
density of SERCA2a on the SR within L-SR junctions to be equivalent to that
previously predicted for PM-SR junctions
as approximately $6250/\mum^2$ \cite{fameli_2007}.
Other input data for the model are the estimates of lysosomal volume
and of the $\Cacon_{\rm lys}$, from which we
calculated the actual number of ions in
the lysosome prior to the beginning of \ca\ release (see previous
section). This, in turn, was used in the extrapolation of the
TPC2 complex \ca\ release rate as a function of time, as follows.

A recent electrophysiological study of the \ca\ conductance of
TPC2 signaling complex
provides valuable information regarding its biophysical properties
and, importantly,
highlights the fact that, for a given relatively low activating concentration
of NAADP (10 nM, as in figure {\ref{Fig_1}), the
channel open probability appears to depend on $\Cacon_{\rm lys}$
(this is likely due to a partial neutralization of the electrochemical
potential across the lysosomal membrane) \cite{Pitt_2010}.
We used the \ca\ conductivity measured in
\cite{Pitt_2010} and values of the lysosome membrane
potential (\cite{Koivusalo_2011}) to calculate the maximal \ca\ current,
$I_{\rm max}$, as $2.4\times 10^6$~ions/s.
We then applied a
weighted quadratic fit to the channel's open probability $(P_{\rm o})$
as a function of the
$\Cacon_{\rm lys}$ data in \cite{Pitt_2010}
with constraints
that the $P_{\rm o}$ would tend to zero at  $\Cacon_{\rm lys}=80\;\muM$,
based on
the observation that below  $\Cacon_{\rm lys}=100\;\muM$
essentially no single channel
openings were observed \cite{Pitt_2010}. Another constraint for the fit
was that the curve be within the standard deviation value
of $P_{\rm o}$ at the highest reported  $\Cacon_{\rm lys}=1\;\mM$.
From the quadratic fit, we then obtained our own
$P_{\rm o}$-vs-$\Cacon_{\rm lys}$ table (plotted in figure \ref{Fig_5}A) and
assumed that at the
beginning of \ca\ release, the \ca\ current would be $P_{\rm o, max}\times
I_{\rm max}$ at the maximal luminal concentration until the luminal
concentration decreased to
the next point in the $P_{\rm o}$-vs-$\Cacon_{\rm lys}$ relationship.
At this point, the release rate decreases to a new (lower) $P_{\rm o}\times
I_{\rm max}$ until the luminal concentration reaches the next lower point in
the table,
and so on until $P_{\rm o}=0$ at $\Cacon_{\rm lys}=80\;\muM$. The \ca\
release rate as a function of time obtained in this manner is shown in figure
\ref{Fig_5}B.
\bfig[!ht]
\begin{center}
\includegraphics*[scale=.5, viewport=0 170 822 455]{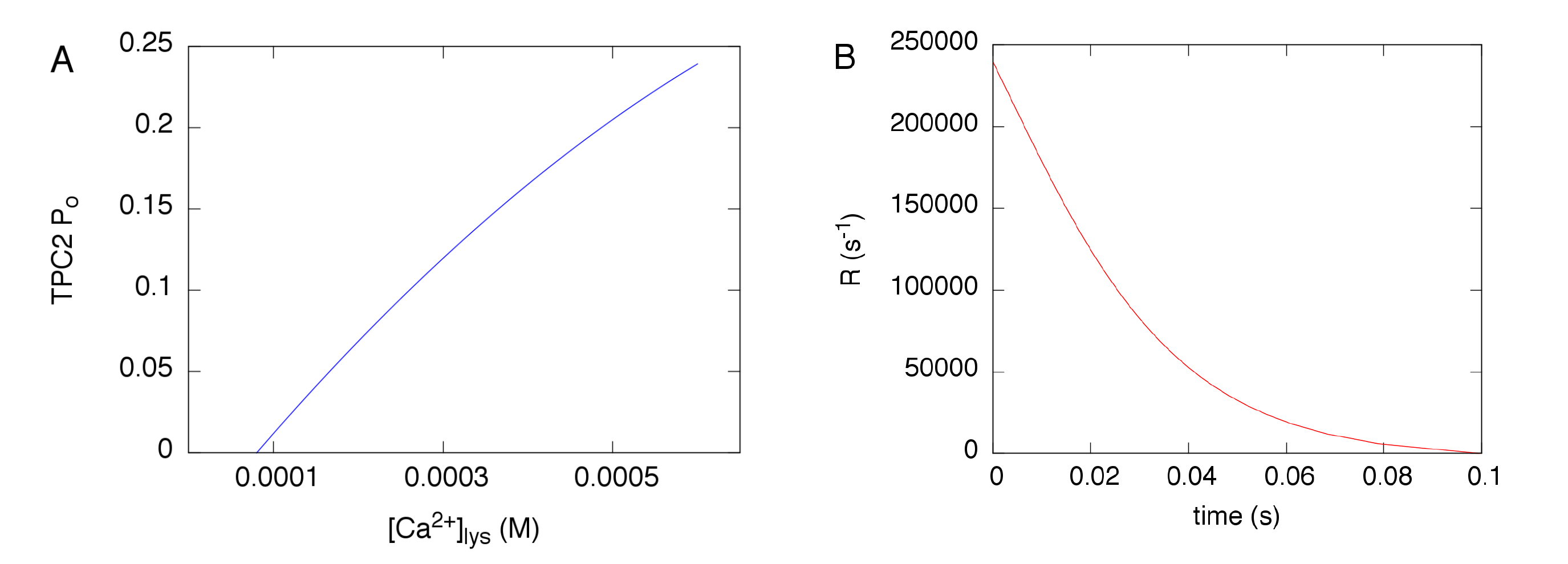}
\end{center}
\caption{\textbf{A,} open probability $P_{\rm o}$ of the TPC2-dependent \ca\
conductance reproduced from a
quadratic fit to the data reported in \cite{Pitt_2010}.
\textbf{B,} \ca\ release rate calculated as explained in the text, based
on the $P_{\rm o}$ in A.
}\label{Fig_5}
\endfig

To implement an approximated SERCA pump action, we used a simplified version of a
multi-state kinetic model developed in \cite{Lauger_1991}, which was
further informed by
more recent isoform-specific studies \cite{Dode_2003,Raeymaekers_2011}.
In particular, since we are
interested in simulating the SERCA2a \ca\ uptake only in terms of its
influence on the shaping of the junctional \cacon\ transient, we have opted
not to implement the steps of the multistate model that deal with the \ca\
unbinding from the SERCA on the SR luminal side of the pump.
In brief, the reactions SERCA2a undergo are: (1)
binding/unbinding of the first \ca; (2) binding/unbinding of the second \ca.

The \ca\ diffusivity was obtained from studies,
in which it was concluded that given the known kinetics of typical \ca\
buffers, the range of free \ca\ after it enters a cell can be up to 200
nm. Therefore, due to the nano-scale of our system, the trajectories of
\ca\ released by TPC2
signaling complexes in the simulations are governed by the measured
diffusivity of free \ca
($2.23\times 10^{-10}\;{\rm m}^2/{\rm s}$;\cite{allbritton_1992, allbritton_1993}).
Once \ca\ are buffered they acquire the measured diffusivity of the
buffers ($8.4\times 10^{-11}\;{\rm m}^2/{\rm s}$;\cite{Keller_2008}).

{We summarize the quantitative model input data in table
\ref{input_data}.}
\begin{table}
\begin{center}
\caption{
\bf{Quantitative model input data}}\label{input_data}
        \begin{tabular}{c|c|c}\hline\hline
        quantity & value & source \\\hline
        $\Delta\Cacon_{\rm shoulder}$ & 300 nM & \cite{boittin_2002} 
	and this article\\
        $\Cacon_{\rm lys}$ & 400--600 $\muM$ & \cite{christensen_2002}\\
        lysosomal average radius & 180 nm & this article, figure \ref{Fig_3}\\
        (cylindrical) SMC radius & 6 $\mum$ & \cite{Somlyo_1975}\\
        SMC length & 130 $\mum$ & \cite{Somlyo_1975}\\
        SMC cytosolic volume & $2.4\times 10^{-12}$ L & this article\\
        TPC2 complexes per lysosome & 20 & \cite{Schieder_2010a, Pitt_2010} and this article\\
        TPC2 complex \ca\ release rate & variable & this article, figure \ref{Fig_6} \\
        SERCA2a surface density & $6250/\mum^2$ & \cite{fameli_2007}\\
        SERCA2a first \ca\ binding (unbinding) rate & $5\times 10^8\,\M^{-1}{\rm	s}^{-1}$ ($60\,{\rm s}^{-1}$) & \cite{Lauger_1991,Dode_2003,Raeymaekers_2011}\\
        SERCA2a second \ca\ binding (unbinding) rate & $4\times 10^8\,\M^{-1}{\rm s}^{-1}$ ($60\,{\rm s}^{-1}$) & \cite{Lauger_1991,Dode_2003,Raeymaekers_2011}\\
        free \ca\ diffusivity, $D_{\rm free}$ & $2.23\times 10^{-10}\;{\rm m}^2/{\rm s} $ & \cite{allbritton_1992, allbritton_1993}\\
        \ca\ buffer diffusivity, $D_{\rm buffer}$ & $8.4\times 10^{-11}\;{\rm m}^2/{\rm s} $ & \cite{Keller_2008}\\\hline\hline
        \end{tabular}
\begin{flushleft}
Summary of input parameters used in various phases of the quantitative
model, and references to the origin of their values.
\end{flushleft}
\end{center}
\end{table}

Using MCell as a stochastic particle
simulator, we ran a number of simulations to represent the NAADP-mediated
\ca\ release that is supposed to occur
at L-SR junctions in PASM experiments. Released \ca\
is assumed to undergo Brownian motion in the
surrounding space, including the L-SR nanospace. SERCA2a placed on the
neighbouring SR surface may capture
\ca\ according to our approximation of their known multistate model.
To determine how this regenerated cellular environment can shape a \ca\
transient we ``measured'' the junctional \cacon\ by counting the ions
within the L-SR volume at any given time and dividing the number by the
volume.
The snapshots in figure \ref{Fig_6} are part of the visual output
of this phase of the work.
The data to determine the simulated \ca\ transients in the L-SR
junctions were collected in a measuring volume placed between the lysosomal
and SR membranes in the virtual L-SR junction (rust-coloured box in
figure \ref{Fig_6}B).

Previous experimental findings about VSMC PM-SR junctions indicate that
disruption of nanojunctions can have profound consequences on \ca\
signaling properties of the cells. For example, when calyculin A was
used to separate superficial SR portions from the plasmalemma of the
rabbit inferior vena cava, it was observed that \caconin\ oscillations
would cease \cite{Lee_2005}. This was later corroborated by a quantitative
model of the PM-SR junction's role in the refilling of SR \ca\ during
oscillations \cite{fameli_2007}. In another study by the van Breemen
laboratory, it
was observed that the mitochondria-SR junctions of airway SMC displayed a
variable average width as a function of the state of rest or activation
of the cell \cite{Dai_2005}. It makes sense then to
study the effect of changes in the junctional geometry on the
transients generated by our simulations. Therefore,
we ran several simulations, in which the separation
between the lysosomal and SR membranes was increased from 10 nm to 100
nm in steps of 10 nm. In figure \ref{Fig_7}A, we report three
sample transients from this set
of simulations obtained using three different junctional 
{membrane separations}, as
indicated in the inset legend. The value of
each of the points graphed in this plot is the
average value of 100 simulations, in each of which the random number
generator
within MCell is initiated with a different seed (see Materials and Methods).
We report representative error bars as
$3\times$ the standard error, to convey the
$> 99\%$ confidence interval of the data.
As one would expect, the transient {nanojunctional (NJ)} \ca\ concentration, 
\caconj, decreases in magnitude, as the junctional L-SR membrane-membrane
separation increases, simply because the
released \ca\ has a larger junctional volume available over which to
spread. However, it is important to make a quantitative comparison between
this change in \caconj\ and the \caconin\ requirements to activate the
putative RyR3 population of the junctional SR. This can be attained by
analyzing the time scale of both the recorded and simulated \ca\ signals.
\subsubsection{Reconciling the temporal scales in
simulation and experimental results}
In the final step of the development of our model, we analyzed the
relationship between the time scale of the \caconj\ transients
resulting from our model (figure \ref{Fig_7})
and that of the observed \ca\ signal in figure \ref{Fig_1}.
\bfig[!ht]
\begin{center}
\includegraphics*[scale=.6, viewport=0 170 822 455]{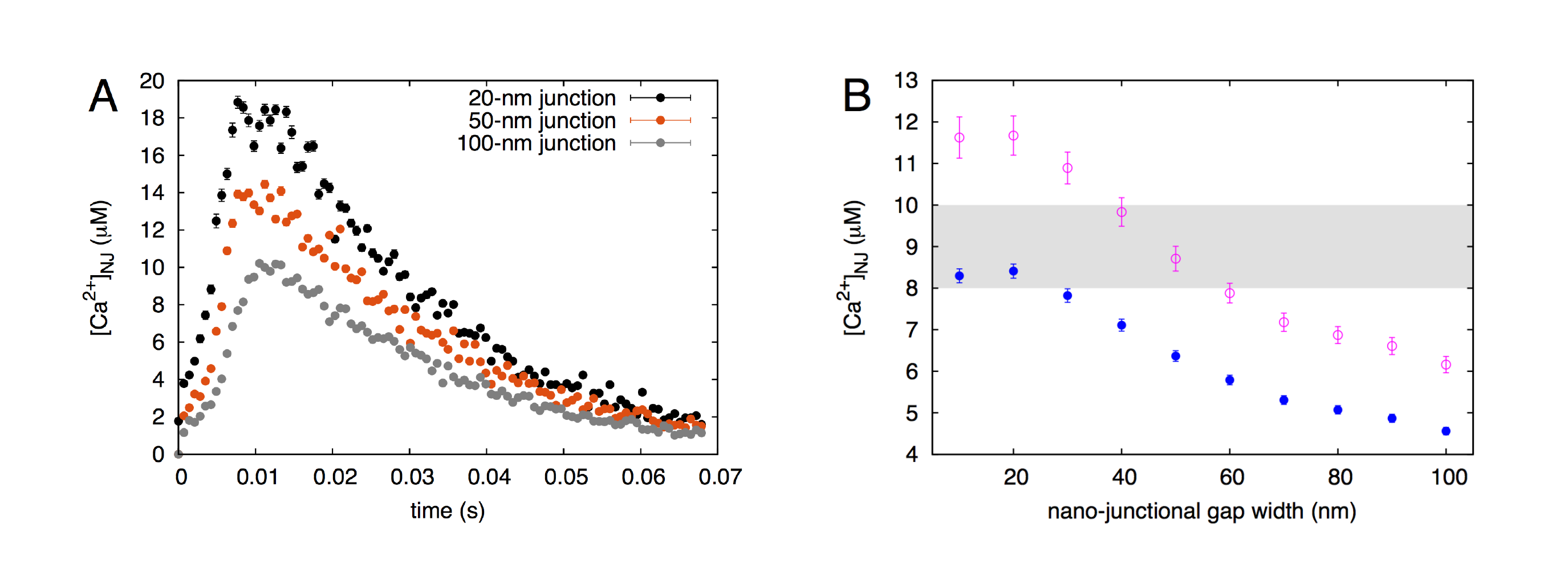}
\end{center}
\caption{\textbf{A,} calculated nanojunctional \cacon\ transient,
$\Caconj$, ``measured'' inside the volume of the recreated L-SR
nanojunction shown in figure \ref{Fig_6}. To show the effect of changes in
the
junctional geometry, we report three transients calculated using
different junctional widths of 20, 50 and 100 nm.
\textbf{B,} $\Caconj$ \textit{vs} width of junction, concentration
values are temporal averages of the transients as in panel A, calculated
over
an interval of 0.065 s (solid circles) and {0.038 s} (empty circles); see
text
for explanation.
The shaded area indicates the approximate threshold values for CICR at
RyR3s \cite{Takeshima_1995}.
}\label{Fig_7}
\endfig

{Let us note that the typical duration of the simulated transients, $t_{\rm
transient}$, is about 0.06 s and recall that these represent
$\Delta\Cacon$ above the resting \cacon.
On the other hand, the build up to the
maximum value of $\Delta\Cacon_{\rm shoulder}$ takes about~5~s (we refer to
this time as $t_{\rm shoulder}$; figure \ref{Fig_1}). 
This interval is about two orders of magnitude larger than the duration of the
simulated individual transients (figure \ref{Fig_7}A).
One way to reconcile the hypothesis that
L-SR junctions are at the base of the observed NAADP-induced \ca\ signals such
as the ones in figure \ref{Fig_1} and in particular that the signal shoulder
emerges from lysosomal \ca\ release at L-SR junctions, is to bring forward
the assumption that the signal shoulder may be the result of a sequential summation
effect over many L-SR junctions, each working according to an all-or-none
mechanism of \ca\ release, and
that the ``firing'' of one junction causes a cascading effect across the
set of junctions that produce the shoulder. Then the duration of the
shoulder upstroke can be expressed as:
\begin{equation}
t_{\rm shoulder}=N_{\rm NJ}\times t_{\rm transient}\label{t_shoulder}
\end{equation}
where $N_{\rm NJ}$ is the number of L-SR nanojunctions that yield
$\Delta\Cacon_{\rm shoulder}$.
Since we calculated above that \ca\ release from $N_{\rm NJ}\approx$~60--100
lysosomes would be necessary to produce the observed signal shoulder
magnitude and have shown in the previous section that such release would need to
take place at L-SR junctions,
we can  estimate $t_{\rm transient}$, the duration the
\caconj\ transient, by reversing equation (\ref{t_shoulder}) and
obtaining $t_{\rm transient}=$0.05--0.08~s.
It is noteworthy that
this range of values is obtained in a manner completely independent of our
simulation results, which yielded a similar range of values.}

To gain quantitative insight into the effects of varying the junctional
width,
we then calculated the temporal average of the $\Caconj$
over $t_{\rm transient}$ ({using the middle value of the range} 
calculated via equation (\ref{t_shoulder}))
and graphed it as a function of the junctional width. The result of
this analysis is reported in figure \ref{Fig_7}B (solid blue circles).
{As we have anticipated at the end of the previous section, the
decrease in magnitude of these data is to be expected, however}
in this plot we also
indicated the range of \caconin\
values (shaded area) over which maximum SR \ca\ release via RyR3
is reported to occur in skeletal muscle \cite{Takeshima_1995}. This
comparison underscores the important constraint played by the width of
the L-SR junctions and indicates that, unless the membrane separation is
kept below about 30 nm, it is not possible for \caconj\ to breach the
threshold for RyR3 \ca\ release. Let us also point
out that the
\caconj\ data in figure \ref{Fig_7}B would shift upward, toward
concentration
values that would make the junction more prone to promote RyR3 release, if
the
temporal average were taken over a shorter transient time, $t^{\prime}_{\rm
transient}<t_{\rm transient}$ around the \caconj\ peak.
However, in that case equation (\ref{t_shoulder}) indicates that a larger
$N_{\rm NJ}$ (than {80, picked as the middle of the 60--100 range}) 
would have to
contribute to the signal summation that results in a 5-second $t_{\rm
shoulder}$.
Interestingly, this possibility agrees with our lysosome count from TEM
images (130 lysosomes/cell) and with the argument of natural
redundancy we contemplated to explain the discrepancy between the
calculated and observed lysosome number estimates.
As an exercise we have recalculated the
\caconj\ time-averaged over $t^{\prime}_{\rm transient}$ obtained
using 130 lysosomes in
equation (\ref{t_shoulder}) (empty purple circles in figure \ref{Fig_7}B),
and this indeed shows that RyR3 \ca\ release threshold would be cleared
more readily.
These observations suggest that activation of RyR3s in the L-SR junctions
not only depends on the concentration of \ca\
near them, but also on their exposure time to this concentration.

\section{Discussion}
\subsection{Inter-organellar nanospaces}
We have recently introduced the concept of the ``pan-junctional SR'', 
which states that \ca\ release and uptake at a
family of specific nanojunctions connected by a continuous but variable SR
lumen integrates cellular control over multiple functions
\cite{breemen_2013}. 
The {lysosome-sarco/endoplasmic reticulum (L-S/ER)}
junction is the most current junction to be considered in this context and
exhibits perhaps the highest degree of plasticity of the family of
nanojunctions of the SR; 
the mechanism and function of lysosomal \ca\ signaling is currently hotly
debated \cite{Lam_2013}. 

By means of a thorough ultrastructural study in rat pulmonary artery smooth
muscle, we have observed and characterized L-SR nano-junctions, 
which had been previously hypothesized on the basis of
optical measurements of \ca\ signals and optical immunocytochemistry
\cite{kinnear_2004, kinnear_2008}. 
Our observations corroborate the previously
reported finding (in \cite{kinnear_2004} and \cite{boittin_2002})
that lysosomes in PASMCs tend to cluster in the perinuclear region, as
is evident in our micrographs (e.g., figure
\ref{Fig_2}A). We find that L-SR junctions are on average 15 nm in
width (equivalent to our preliminary reports \cite{edinburgh_2011, bps_2013} 
and to recent observations in cultured fibroblasts
\cite{kilpatrick_2012}) and extend approximately 300 nm in
lateral dimensions, thereby involving about 15\% of the lysosomal membrane
(table \ref{l_sr_params}).

\subsection{Mechanism of NAADP $\boldsymbol{\Ca}$ signaling}
In an effort to achieve quantitative understanding of the phenomenon of
NAADP-mediated \ca\ transients and verify the proposal
that these may be generated in L-SR junctions \cite{kinnear_2004, 
boittin_2002}, we focused on one of the prominent features of these \ca\
signals, namely the localized \ca\ bursts that precede the propagating \ca\
wave (figure \ref{Fig_1}), which we refer to as the signal
shoulder ($\Delta\Cacon_{\rm shoulder}$). 
In the first instance, we have estimated the potential contribution of 
local bursts of \ca\ release from individual lysosomes 
to the elevation of global \caconin\ observed during this shoulder
in the experimental records. From this, and using the dimensions of a 
typical smooth muscle cell, we calculated
that {60--100} lysosomes would be required to cause an elevation of
comparable magnitude to the signal shoulder. This is
lower than the estimate for the total number of lysosomes per cell,
{130}, we obtained from our ultrastructural study, 
but comparable in order of magnitude. 
We have already mentioned above a number of factors in favour of observing a greater
number of lysosomes/cell than the estimated number
required to generate the signal shoulder. 
In addition, several other elements contribute a degree of
uncertainty to those estimates, such that their discrepancy may not be as
large. We need to consider that $P_{\rm o}$ data for the \ca\
conductance associated with the TPC2 signaling
complex published in
\cite{Pitt_2010}, on which we
based our TPC2 rate table, show some variability according to the
standard deviation bars, which, in turn, implies an uncertainty 
in the interpolated  $P_{\rm o}$
(figure \ref{Fig_6}A). Moreover, we cannot exclude the possibility that
these data reflect a contribution from multiple channels 
($NP_{\rm o}$) rather than a purely single channel $P_{\rm o}$.
Lastly, the standard deviation of the simulated transient (only the standard error
is shown in figure \ref{Fig_7}A) and the variability in the experimental 
determination of the \cacon\ sensitivity of RyR may allow for some 
uncertainty in the estimated number of nanojunctions. 

To take this study further and understand whether the observed L-SR junctions
could give rise to \caconin\ transients of appropriate magnitude to trigger
\ca\ release from RyR3 channels on the junctional SR, we developed a
quantitative stochastic model of \ca\  dynamics in the junctional
nanospaces. We have
previously published a simplistic version of this model, which
nonetheless captured the essential features of the problem and yielded an
indication that such \ca\ transients in the L-SR
junctions would be possible \cite{breemen_2013}. 
However, the simulated transients we obtained displayed
unphysiological features, such as an abrupt onset and decay. We show here
that this was largely
due to lack of a faithful representation of the open probability for the
\ca\ conductance of the TPC2 signaling complex. We
have now combined experimental information on the biophysical properties
of conductance and open probability (from \cite{Pitt_2010}) and on the luminal 
\cacon\ of
the lysosomes (from \cite{christensen_2002}) 
to implement a more realistic \ca\ release rate model in
the simulations. As a consequence, we are able to output a 
physiologically meaningful junctional transient profile (figure
\ref{Fig_7}A), and observe that the
$\Caconj$ transients generated within our model junctions---in turn, 
based on those
observed in our TEM images---reach peaks of about $20\;\muM$. While we are
still not able to measure these transients in individual 
L-SR junctions
(peri-lysosomal \ca\ probes are only recently becoming available---see for example
\cite{McCue_2013}---and so far have not been used in smooth muscle cells,
vascular or otherwise), it is worth noticing 
that the values we find are comparable to those recently 
measured in so-called \ca\ hot spots in the mitochondria-ER junctions of
neonatal ventricular cardiomyocytes 
(rat culture) \cite{Drago_2012} and in RBL-2H3 and
H9c2 cells earlier \cite{Csordas_2010}.
These results 
therefore suggest that the hypothesis
presented for the role of L-SR junctions in cellular \ca\ signaling 
is certainly plausible. 

\subsection{Lysosomal trigger zone}
Although our simulations were successful,
comparison of the \ca\ transients in figures \ref{Fig_1}
and \ref{Fig_7} not surprisingly reveals a striking difference 
between the simulated 
\ca\ kinetics of a single L-SR junction and those of the whole cell.
This illustrates important aspects related to the concept of lysosome-SR
``trigger zone'' previously introduced by one of us (AME, \cite{kinnear_2004}) 
{and recently proposed as a facilitating factor in lysosomal-ER  
signaling in reverse, whereby \ca\ release from ER compartments enables 
NAADP-mediated activation of acidic organelles \cite{Morgan_2013}}. 

{Briefly, this concept 
suggests that in PSMCs clusters of lysosomes and
closely apposed SR regions containing sets of RyRs may act together to form 
specialized intracellular compartments that transform NAADP-stimulated localized
lysosomal \ca\ release into cell-wide \ca\ signals via RyR-supported CICR.}
If the firing
of one junction were sufficient for subsequent initiation of the CICR across
the entire SR, the experimentally observed threshold (figure
\ref{Fig_1}) would be
much lower and the rate of junctional coupling by CICR 
faster. 
In other words, due to
the high \caconin\ threshold of about 10 $\muM$ for \ca\ activation of RyR3, 
CICR engendered by a single L-SR junction is likely to die out, unless
reinforced by a process of quantal releases 
by other L-SR junctions within a cluster \cite{Zhu_2010}. 
Therefore it seems more likely that multiple L-SR
junctions work in concert in a
process characterized by both additive and regenerative elements 
to provide the necessary threshold and margin of safety
required to ensure the propagation by CICR of the global wave via 
the more widely distributed RyR2 along extra-junctional SR, 
once the \ca\ release wave escapes an RyR3-enriched SR region, 
as previously suggested in \cite{kinnear_2008}. 

In this respect, our results also allowed us to establish a time interval for the
duration of the transient ($t_{\rm transient}$) 
that would be compatible with the hypothesis that
sequential summation of 
all-or-none lysosomal \ca\ release events from a set of individual 
L-SR junctions
is responsible for generating $\Delta\Cacon_{\rm shoulder}$. Remarkably, this
value is comparable to the average duration of the simulated transients, which was
determined on the basis of ultrastructural details 
and completely independently of the summation effect hypothesis.
The assumption of summation may imply that the role of the NAADP as a stimulus
is limited to the initial lysosomal \ca\ release, while the recruitment of
subsequent L-SR junctions may be governed by further lysosomal calcium
release, by SR \ca\ release via \ca\ activated junctional RyR3, 
and/or propagation and combination of calcium signals via
inter-junctional clusters of RyR3. Therefore, from our model 
we envision the intriguing possibility of  
an important regulatory role of lysosomal \ca\ content by SR \ca, in such a
way that
summation of calcium signals at
multiple junctional complexes may give rise to the shoulder. 
While only further experiments can verify this,
the quantitative
corroboration provided by our findings sets on firmer ground the conclusion
from earlier studies that \ca\ bursts from lysosomes are indeed responsible
for initiating the first phase of a cell wide \ca\ wave via L-SR
junctions.

\subsection{Disruption of nanojunctions, plasticity and pathology}
Carrying the signal time-scale analysis further and
using the calculated $t_{\rm transient}$ to take a temporal average
of the simulated \caconj\ at different junctional widths, we find that
above a width of about 30 nm  the transients would be
unable to trigger \ca\ release from the RyR3s (figure \ref{Fig_7}B).
Thus it is possible that heterogeneity and plasticity are controlled by
a variable width of the junctional nanospace.
For example, in atrial myocytes it has been proposed that NAADP
evokes \ca\
release from an acidic store, which enhances general SR \ca\ release by
increasing SR
\ca\ load and activating RyR sites
\cite{Collins_2011}. 
This functional variant may be
provided by either:
(1) An increase in junctional distance such that $\Cacon_{\rm
NJ}$ is
insufficient to breach the threshold for activation of RyR2, yet
sufficient
to allow for increases in luminal \ca\ load of the SR via
apposing SERCA2
clusters; or (2) L-SR junctions in cardiac muscle formed between
lysosome
membranes and closely apposed regions of the SR which possess
dense SERCA2
clusters and are devoid of RyR2.  Further ultrastructural studies
on cardiac
muscle and other cell types may therefore provide a greater
understanding of
how L-SR junctions may have evolved to provide for cell-specific
modalities
within the calcium signaling machinery.

{Under metabolic stresses, such as hypoxia, lysosomal pathways have been
proposed to provide for autophagic glycogen metabolism via acid maltase in
support of energy supply, and significant levels of protein breakdown during
more prolonged metabolic stress. Moreover, although controversial, in some
cell types it has also been suggested that lysosomes may contribute to the
energy supply by providing free fatty acids for beta-oxidation by mitochondria
\cite{Kovsan_2010}.  It may be significant, therefore, that TPC2 gating and
thus autophagy may be modulated by mTOR \cite{Cang_2013, Lu_2013}. 
This may well speak to further
roles for lysosomal calcium signaling and L-SR junctional plasticity during
hypoxic pulmonary vasoconstriction and even during the development of hypoxia
pulmonary hypertension \cite{Evans_2009}, 
not least because AMP-activated protein kinase has
been shown to modulate autophagy through the phosphorylation and inhibition of
mTOR (see for example, \cite{Park_2013}).}

The process of autophagy serves not only to regulate programmed
cell death, but also to recycle organelles, such as mitochondria, through a
process of degradation involving lysosomal hydrolases \cite{Gozuacik_2004}. 
For this to occur L-SR junctions would likely be disrupted
and thus select for local rather than global \ca\ signals in order to
facilitate fusion events among lysosomes, endosomes, autophagosomes and
amphisomes \cite{Klionsky_2008}, which follows from the fact that \ca\
plays a pivotal role in vesicle trafficking and fusion \cite{Luzio_2000}. 
Active non-synchronous movements of
TPC2-expressing vesicles have been detected in live-cell imaging experiments
using GFP-tagged proteins \cite{Calcraft_2009}. Thus, lysosomal \ca\
signaling may modulate plasticity in the manner required at all stages of
multiple membrane fusion events that are dependent on \ca\ for the effective
formation of the SNARE complexes \cite{Luzio_2000}. In short, transport
of proteins between lysosomes, Golgi apparatus, and plasma membrane via
lysosomes and endosomes may be coordinated at different stages by both global
\ca\ signals and spatially restricted \ca\ release from acidic stores in a
manner that is in some way determined by L-SR junctional integrity.
 
Loss of integrity of L-SR junctions may also contribute to disease, given that
lysosomal \ca\ release both depletes luminal \ca\ and causes intraluminal
alkalinization \cite{Morgan_2007}. Therefore, disruption of L-SR junctions
may modulate the activity of pH-sensitive hydrolytic lysosomal enzymes, such
as glucocerebrosidase and acid sphingomyelinase, which exhibit a marked loss
of function at pH $>5$ \cite{Weely_1993, Schneider_1967} that could
lead to accumulation of macromolecules such as glucocerebroside and
sphigomylin. As mentioned previously, dysfunction of these 
enzyme systems consequent to L-SR
junctional abnormalities could also contribute to 
pathologies associated with subclasses of lysosome storage disease
such as Niemann-Pick disease type C1 \cite{lloyd-evans_2008, Tassoni_1991},
Pompe and Gaucher \cite{Noori_2002, Jmoudiak_2005}, which may include hepatic
portal \cite{Tassoni_1991}, or pulmonary hypertension \cite{Noori_2002,
Jmoudiak_2005}, dysfunctions in cholesterol
trafficking \cite{Carstea_1997} 
and consequent increases in plasma cholesterol levels, vascular
lesion formation, atherosclerosis/thrombosis and medial degradation
\cite{Tassoni_1991, Ron_2008}. 

\subsection{Conclusion}
We have determined that L-SR junctions, about 15 nm in
width and extending to $\approx 300$ nm in lateral dimensions
are a regularly occurring feature in rat pulmonary arterial
smooth muscle cells, in which L-SR junctions were first proposed.
The present study provides a mechanistic basis for the observed 
NAADP-induced \ca\ signals  
and strong support for the proposal
that these \ca\ signals are generated at L-SR junctions. 
Even within the variability of our recorded values of L-SR junctional 
widths and extension, 
our results suggest that localized \cacon\ transients due to junctional
\ca\ release can without fail reach values required to breach the threshold for
CICR from junctional RyR3s. Perhaps most significantly, 
disruption of the nanojunctions decreases
$\Caconj$ below the value for CICR via junctional RyR3s. 
Therefore, consistent with previous studies on the PM-SR membrane 
\cite{fameli_2007}, 
we have established that L-SR junctions are required to allow such signals to
be generated and that there is a 30--50 nm
limit on junctional width, above which there is loss of junctional integrity
and inadequate control of ion
movements within the junctional space. 
This suggests that the observed L-SR
junctions in PAMSCs are not only capable of delivering localized
\ca\ bursts of the required magnitude, but are also necessary if lysosomes
are to fulfill this identified role in \ca\ signaling.
In other words,  L-SR nanojunctions are
a necessary and sufficient condition for generating local \ca\ bursts
essential for NAADP-induced \ca\ waves.
Most importantly, however, this study demonstrates the importance of
junctional architecture on the nanoscale to the capacity for coupling across
cytoplasmic nanospaces, tight regulation of ion transport and thus signal
transduction. In turn, this highlights the possibility that alterations in
the dimensions and architecture of intracellular nanojunctions lead to
cell dysfunction and hence disease.

\section{Materials and methods}
All the experiments and procedures were carried out
in accordance with the guidelines of the University of
British Columbia Animal Care Committee and in accordance with the United
Kingdom Animals (Scientific Procedures) Act 1986.
\subsection{Electron microscopy}\label{em_methods}
Male Wistar rats weighing 220--250 g were anesthetized with 3 mL of Thiotal. 
The thoracic cavity was
opened and flooded with warm physiological saline solution
(PSS) containing 145 mM NaCl, 4 mM KCl, 1 mM MgCl$_2$, 10 mM HEPES,
0.05 mM CaCl$_2$, and 10 mM glucose (pH 7.4). After extraction of the
heart and lungs and their placement in HEPES buffer, rings from the primary
and secondary branches of the pulmonary artery were dissected and 
immediately immersed in fixative solution. 
The primary fixative solution contained 2.5\% glutaraldehyde in
0.1 M sodium cacodylate buffer at room temperature.
The artery rings were then washed three times in 0.1 M sodium
cacodylate (30 min in total). In the process of secondary fixation, the
tissue rings were fixed with 1\% OsO$_4$ in 0.1 M sodium
cacodylate buffer for 1 h followed by three 10-minute 
washes with distilled water and \textit{en bloc} 
staining with 2\% uranyl acetate.
The samples were then dehydrated in increasing concentrations of ethanol 
(25, 50, 75, 80, 90, and
95\%).
In the
final process of dehydration, the samples underwent 3 washes in 100\%
ethanol. 
The artery rings were then resin-infiltrated in increasing
concentrations (30, 50, and 75\% in ethanol) of a 1:1 mix
of Epon and Spurr's resins. 
The infiltration process was completed by three
passages in 100\% resin. All of the ethanol dehydration and 
resin infiltration steps were
carried out by using a laboratory
microwave oven. 
The blocks were finally resin-embedded in
molds and polymerized overnight in an oven at $60^{\circ}$C. 
\subsubsection{2D imaging}
For standard (2D) electron microscopy imaging, 80-nm sections were cut from
the embedded sample blocks on a Reichert Ultracut-E microtome
using a diamond knife and were collected on
uncoated 100- and 200-mesh copper grids (hexagonal or square meshes). 
The sections were post-stained with
1\% uranyl acetate and Reynolds lead citrate for 12 and 6 minutes,
respectively. Electron micrographs at various magnifications 
were obtained with a Hitachi 7600 transmission electron
microscope at 80 kV. 

{Lysosomes in these images were identified according to their well
established appearance features: they are single lipid bilayer membrane-bound, 
with a granular, more or less uniform luminal matrix 
that is more electron dense than the surrounding
cytosol. Secondary lysosomes may also contain less
granular structures within the finer matrix. 
Moreover, lysosomes are normally distinguishable from endosomes by their
larger size, hence we set a threshold 
``diameter'' of $>200$~nm for acceptance of a lysosome, below
which all vesicles were excluded.}
\subsubsection{Tomography (3D imaging)}
To obtain electron microscopic tomograms, we cut 200-nm-thick sections
from the same sample blocks used for standard imaging. The sections were
then collected on Formvar coated slot copper grids and post-stained
with
1\% uranyl acetate and Reynolds lead citrate for 20 and 10 minutes,
respectively. We surveyed the sample sections using a FEI Tecnai G2
200 kV transmission electron microscope and identified
regions of interest containing L-SR junctions. We then acquired tomograms 
of several of those regions 
by taking 2D scans through the sample every 5$^{\circ}$ 
of inclination as it was
tilted between 
$-60^{\circ}$ and $+60^{\circ}$ with respect to horizontal. The scans
were reconstructed with the Inspect3D software tools 
and structures of interest, primarily lysosomal and SR
membranes in the same cellular neighbourhood, were traced out using
Amira software.
\subsection{Image and data analysis}
The images of the samples were further processed
using GIMP (GNU Imaging Manipulation Program, open source, available at
gimp.org) to enhance
membrane contrast in the interest of improving our characterization of
the L-SR junctions. 

The SR and lysosomal
membranes were outlined, highlighted and measured in pixels using the
Inkscape software (open source, available at inkscape.org), 
converting the pixel measurements to nm using the scale bar
in the recorded micrographs. 
By modifying the Inkscape script for measuring lengths, we were able to
output the measurements directly into a text file, which we used to
produce the histograms in Figure \ref{Fig_3}B--D.
We used the package Gnuplot (open source, available at
gnuplot.info) to produce 
the histograms and plots presented herein. 
\subsection{$\boldsymbol{\Ca}$ imaging}
Single arterial smooth muscle cells were isolated from second-order branches
of the pulmonary artery. Briefly, arteries were dissected out and placed in
low \ca\ solution of the following composition (mM): 124 NaCl, 5 KCl, 1
MgCl$_2$, 0.5 NaH$_2$PO$_4$, 0.5 KH$_2$PO, 15 NaHCO$_3$, 0.16 CaCl$_2$, 
0.5 EDTA, 10 glucose,
10 taurine and 10 Hepes, pH 7.4. After 10 min the arteries were placed in the
same solution containing 0.5 mg/ml papain and 1 mg/ml bovine serum albumin
and kept at 4$^{\circ}$C overnight. 
The following day 0.2 mM 1,4-dithio-DL-threitol was
added to the solution, to activate the protease, and the preparation was
incubated for 1 h at room temperature (22$^{\circ}$C). The tissue was then
washed at 3$^{\circ}$C
in fresh low \ca\ solution without enzymes, and single smooth muscle cells
were isolated by gentle trituration with a fire-polished Pasteur pipette.
Cells were stored in suspension at 4$^{\circ}$C until required.

PASMCs were incubated for 30 min with 5 $\muM$ Fura-2-AM in \ca-free PSS
in an experimental chamber on a Leica DMIRBE inverted microscope and then
superfused with Fura-2 free PSS for at least 30 min prior to experimentation.
Intracellular \ca\ concentration was reported by Fura-2 fluorescence ratio
(F340/F380 excitation; emission 510 nm). Emitted fluorescence was recorded at
22$^{\circ}$C with a sampling frequency of {0.5 Hz}, using a 
Hamamatsu 4880 CCD camera
via a Zeiss Fluar 40$\times$, 1.3 n.a. oil immersion lens and Leica DMIRBE
microscope. Background subtraction was performed on-line. Analysis was done
via Openlab imaging software (Improvision, UK).

NAADP was applied intracellularly in the whole-cell configuration of
the patch-clamp technique, and in current clamp mode ($I = 0$) as described
previously \cite{boittin_2002}. 
The pipette solution contained (in mM): 140 KCl, 10 Hepes, 1
MgCl$_2$ and 5 $\muM$ Fura-2, pH 7.4. The seal resistance, as measured using an
Axopatch 200B amplifier (Axon Instruments, Foster City, CA), was $\ge 3$
G$\Omega$
throughout each experiment. Series resistance and pipette resistance were
$\le$ 10 M$\Omega$ and $\le$ 3 M$\Omega$, respectively. 
All experiments were carried out at room
temperature ($\approx 22^{\circ}$ C).
\subsection{Quantitative modeling}\label{qm_methods}
The main stages of the quantitative modeling approach are: 
\begin{enumerate}
\item the
design of 3D software mesh objects (nets of
interconnected triangles by which surfaces can be represented in
computer graphics) representing
a typical L-SR region, including a
whole lysosome and a portion of neighbouring SR network. These objects
are
built to-scale following the ultrastructural characterization of the
L-SR junctional regions as it results from our electron microscopy
image analysis; this phase was carried out using the ``3D content
creation suite'' Blender (open source, available at blender.org);
\item the positioning of the relevant transporters
on the reconstructed membranes (TPC2 complexes on the lysosome,
SERCA2a and RyR3 on the
SR) according to information gathered from the literature 
on their typical membrane densities
and the implementation of the transporter known kinetics and multistate
models and of the ion
diffusivities (\ca\ and mobile \ca\ buffers); 
\item the simulation of
molecular Brownian motion in the cytosol 
by random walk algorithms; this phase was performed by writing appropriate
code for the
stochastic particle simulator MCell (freely available at mcell.org)
\cite{stiles_1996, stiles_2001, kerr_2008}.  In a nutshell, MCell
reproduces the randomness of the molecular trajectories, of the ion
transporter flickering and of the relevant chemical reactions by
probabilistic algorithms governed by random number generators 
(iterative mathematical algorithms, which produce a random
sequence of numbers once initiated by a given number called seed).
This enables the simulation of a number of microphysiological
processes,
all stochastically different from one another. 
The average outcome of 
the processes, to mimic the instrumental output during
experimental measurements, is obtained by taking the average of a
desired quantity, e.g., \cacon, over a large number of simulations
all initiated with a different seed;
\item the measurement of simulated \cacon\ in
the L-SR junctions from the process of
\ca\ release via TPC2-related signaling complex 
on the lysosome and SR \ca\ uptake by the SERCA2a
pumps, and static as well as dynamic visualization of the simulations; this
stage is part of MCell's output. 
\end{enumerate}

\section*{Acknowledgements}
We are very grateful to Garnet Martens and the University of British
Columbia Bioimaging Facilty for their assistance.
We also acknowledge the help of 
David Walker and Arash Tehrani with TEM image analysis and acquisition.
This research has been enabled by the use of computing resources provided 
by WestGrid and Compute/Calcul Canada (westgrid.ca, computecanada.ca).

\bibliography{lys_SR}

\end{document}